
\magnification=1200
\headline{\ifnum\pageno=1 \nopagenumbers
\else \hss\number \pageno \fi}
\overfullrule=0pt
\footline={\hfil}
\font\boldmath=cmbsy10
\textfont2=\boldmath
\mathchardef\mynabla="0235
\def\bfnabla{{\fam=2 \mynabla}\fam=2}
\font\boldgreek=cmmib10
\textfont9=\boldgreek
\def\bfsigma{{\fam=9 \mysigma}\fam=1}
\mathchardef\mysigma="091B
\centerline{{\bf Coulomb Systems Seen as Critical Systems :}}
\centerline{{\bf Ideal Conductor Boundaries}}\bigskip
\centerline{{\bf B. Jancovici}\footnote{$^1$}{Laboratoire de Physique
Th\'eorique et Hautes
Energies, Universit\'e de Paris-Sud, 91405 Orsay, France (Laboratoire
Associ\'e au Centre National de la Recherche Scientifique - URA 63).
E-mail : janco, tellez @ stat.th.u-psud.fr} {\bf and G. T\'ellez$^{1}$}}

\vskip 2cm
\noindent \underbar{Abstract} \par
\baselineskip=20 pt
The grand potential of a classical Coulomb system has universal
finite-size corrections similar to the ones which occur in the free
energy of a simple critical system : the massless Gaussian field.
Here, the Coulomb system is assumed to be confined by walls made of
an ideal conductor material~; this choice corresponds to simple
(Dirichlet) boundary conditions for the Gaussian field. For a
$d$-dimensional ($d \geq 2$) Coulomb system confined in a slab of
thickness $W$, the grand potential (in units of $k_BT$) per unit area
has the universal term $\Gamma (d/2) \zeta (d)/2^d \pi^{d/2}
W^{d-1}$. For a two-dimensional Coulomb system confined in a disk
of radius $R$, the grand potential (in units of $k_BT$) has the
universal term $(1/6) \ln R$. These results, of general validity,
are checked on two-dimensional solvable models. \vskip 1 truecm
\noindent {\bf KEY WORDS~:} Critical systems~; finite size effects~;
Coulomb systems~; solvable models.

\vskip 2 truecm \noindent LPTHE Orsay 95-10 \par
\noindent March 1995 \par

\vfill \supereject \baselineskip=20pt
\noindent {\bf 1. INTRODUCTION} \vskip 5mm
A classical Coulomb system is a system of charged particles, interacting
through the Coulomb
potential, plus perhaps some short-range interaction~; there may also be a
continuous charged
background. We are interested in equilibrium properties,
 and classical (i.e. non-quantum)
statistical mechanics is used. A Coulomb system may have phase
transitions (at least in the two-dimensional case)~; here we assume the
system to be in a conducting phase. \par

In some geometries, it has already been shown$^{(1,2)}$ that the free energy
(or the grand
potential) of a classical Coulomb system exhibits universal finite-size
corrections, in close
analogy with what happens in critical systems$^{(3-6)}$. In  Coulomb
systems, although the screening effect makes the charge correlations
short-ranged, that same screening effect makes the electric potential
and field correlations long-ranged$^{(7)}$~; the critical-like behavior of
the free energy is related to the existence of these long-ranged
correlations. The electric potential is in some sense the analog of the
order parameter of a critical system. \par

For defining a Coulomb system with boundaries, one has to state the
boundary conditions. In a previous paper$^{(2)}$, we considered the case
when the boundary is a plain hard wall which confines the charges. Such an
assumption does not generate any simple boundary conditions for the
electric potential~: for a given charge distribution, the potential
``leaks out'', taking in general non-zero values outside the Coulomb
system and that spoils the analogy with the order parameter of a critical
system. \par

The purpose of the present paper is to study classical Coulomb systems
with boundaries, in the simpler case of Dirichlet boundary conditions for
the electric potential~: the walls, impenetrable to the particles of the
Coulomb system, are assumed to be made of an ideal conductor material on
which the electric potential vanishes. In general, one must also assume
that there is some wall-particle short-range repulsion which prevents a
particle from collapsing onto its electric image. \par

Two geometries will be considered.  \par

\vskip 5 mm
(a) {\bf Slab}.
In $d$ dimensions ($d \geq 2$), a Coulomb system is confined between two
parallel ideal conductor plates, separated by a distance $W$~; the Coulomb
system extends to infinity in the $d - 1$ other directions. Let $\omega$
be the grand potential per unit area. We shall show that $\omega$ (times
the inverse temperature $\beta$) has the large-$W$ expansion
$$\beta \omega
= AW + B + {C(d) \over W^{d-1}} + \cdots \eqno(1.1)$$

\noindent The first two terms represent, respectively, the bulk and the surface
contributions~; the
coefficients $A$ and $B$ are non-universal. However, the last term of (1.1)
is universal, with a coefficient $C(d)$ depending only on the dimension
$d$~: $$C(d) = {\Gamma \left ( {d \over 2} \right ) \zeta (d) \over 2^d
\pi^{{d \over 2}}} \eqno(1.2)$$

\noindent where $\Gamma$ and $\zeta$ are the gamma function and the Riemann
zeta function. In
particular, $C(2) = \pi/24$. \par

Before we derived the expansion (1.1), we guessed it by analogy with a
similar expansion which is valid for a simple critical system~: the
massless Gaussian field theory. We define the partition function of that
theory as the functional integral on a field $\phi({\bf r})$ $$Z_G = \int
{\cal D} \phi({\bf r}) \exp \left [ - {\beta \over 2\mu_d} \int \phi({\bf
r})(- \Delta) \phi({\bf r}) d{\bf r} \right ] \eqno(1.3)$$

\noindent where $\mu_2 = 2 \pi$, $\mu_3 = 4\pi$, and more generally, for $d >
2$, $\mu_d = (d - 2)
2 \pi^{d/2}/\Gamma (d/2)$. Some ultraviolet regularization is needed for
avoiding divergences.
With Dirichlet boundary conditions for $\phi ({\bf r})$ on two parallel plates,
it was
shown$^{(3)}$ that the free energy associated to $Z_G$, in the case $d = 2$,
has a universal term
$- \pi/24W$. More generally, we shall show that the expansion of the free
energy for the Gaussian
field theory is of the form (1.1), except for a change of sign of the
universal coefficient $C(d)$.
  \par

\vskip 5 mm
(b) {\bf Disk}.
A two-dimensional Coulomb system is confined in a disk of radius $R$. The
boundary circle is
assumed to be an ideal conductor. For that case, we shall show that the grand
potential $\Omega$
has the large-$R$ expansion
$$\beta \Omega = AR^2 + BR + {1 \over 6} \ln R + \cdots \eqno(1.4)$$

\noindent where again the coefficients $A$ and $B$ of the bulk and perimeter
contributions are
non-universal, while the term $(1/6) \ln R$ is a universal finite-size
correction. Here too, we first guessed (1.4) by analogy with a similar
expansion$^{(5, 6)}$ which holds\footnote{$^2$}{More general domain shapes
were considered in refs. 5 and 6.} for the Gaussian field theory, except
for a change of sign of the universal term. \par

In two dimensions, there are exactly solvable models of Coulomb systems~: the
two-component
plasma and the one-component plasma, at some special temperature. The general
expansions (1.1)
and (1.4) can be explicitly checked on these models. \par

The critical system which is related to Coulomb systems is the massless
Gaussian field theory~;
this relation is discussed in Section 2. Section 3 is about the slab geometry,
Section 4 about
the disk geometry. The Conclusion suggests some further research.

\vskip 5mm
\noindent {\bf 2. COULOMB SYSTEMS AND GAUSSIAN FIELD THEORY}\vskip 5mm
There are many signs that there is some connexion between Coulomb systems
 and the Gaussian field
theory. The Coulomb potential is the inverse of the properly normalized
Laplacian $-(1/\mu_d)
\Delta$ which appears in (1.3). If $\phi$ is interpreted as the electric
potential, the Hamiltonian in (1.3)
$$H = {1 \over 2\mu_d} \int \phi({\bf r}) (-\Delta ) \phi ({\bf r})
d{\bf r} = {1 \over 2 \mu_d}
\int \left [ \bfnabla \phi ({\bf r}) \right ]^2 d{\bf r} \eqno(2.1)$$

\noindent is the familiar expression for the Coulomb energy in terms of the
electric field $-
\bfnabla \phi ({\bf r})$. Beyond some microscopic cutoff distance, the
correlations of the electric potential in a Coulomb system are
Gaussian$^{(7)}$ and of the same form as the correlations associated with
(1.3). \par

However the partition function of a Coulomb system is not identical to (1.3),
in particular because
(1.3) involves a functional integral on the potential rather than the
familiar integral on the particle positions. A Gaussian transformation
was made in ref. 1 for relating the Coulomb system to (1.3), but,
although this has been useful for the slab geometry, we have not been
able to deal with the disk geometry by this method. \par

We suggest here another approach, based on the heuristic assumption that the
universal features of
the grand partition function $\Xi_C$ of a conducting Coulomb system are
correctly accounted for by the functional integral expression
$$\Xi_C = \int {\cal D} \rho \exp \left [ - {\beta \over 2} \int d {\bf r}
\ d{\bf r}' \ \rho({\bf r}) \ G({\bf r}, {\bf r}') \ \rho({\bf r}') \right
] \eqno(2.2)$$

\noindent where $\rho ({\bf r})$ is a charge density and $G({\bf r}, {\bf
r}')$ the Coulomb interaction solution of
$$\Delta G({\bf r}, {\bf r}') = - \mu_d \ \delta ({\bf r} - {\bf r}')
\eqno(2.3)$$

\noindent with the appropriate (here Dirichlet) boundary conditions. In
(2.2), the non-universal particle structure of the Coulomb system is
already disregarded, and some cutoff prescription has to be made.
Performing now a change of function from the charge density $\rho ({\bf
r})$ to the electric potential $\phi ({\bf r})$, which are related to
each other by
$$\Delta \phi ({\bf r}) = - \mu_d \ \rho ({\bf r})$$

\noindent we obtain
$$\Xi_C = {D \rho \over D \phi} Z_G \eqno(2.4)$$

\noindent where $Z_G$ is the Gaussian partition function (1.3) and $D \rho /D
\phi$ the Jacobian
of the transformation from $\rho$ to $\phi$. Since
$$Z_G = \left [ \det \left ( - {1 \over \mu_d} \Delta \right ) \right ]^{-{1
\over 2}}$$

\noindent and
$${D \rho \over D \phi} = \det \left ( - {1 \over \mu_d} \Delta \right
)$$

\noindent one obtains, from (2.4), $\Xi_C = 1/Z_G$ and
$$\ln \Xi_C = - \ln \ Z_G \ \ \ . \eqno(2.5)$$

\noindent Thus, $\ln \Xi_C$ and $\ln Z_G$ have the same universal
term, except for its sign. The change of sign comes from the Jacobian
$D\rho /D\phi$ in (2.4), i.e. the replacement of the charge density by
the electric potential, as already explained in a different language in
ref. 2.

\vskip 5mm \noindent {\bf 3. {SLAB}} \par \vskip 5mm
We consider a slab in $d$ dimensions. Let us write the $d$-dimensional
 position vector as ${\bf r} =
({\bf x}, y)$, where $y$ is the $d$th component and ${\bf x}$ stands for the
$d-1$ other components.
The Coulomb fluid fills the slab between $y = 0$ and $y = W$, where there are
ideal conductor walls.
\par

In this Section, we derive the large-$W$ expansion (1.1) in general, in
two different ways, and we check its validity on two two-dimensional
solvable models. We make a comparison with the case of plain hard walls.
\par \vskip 5 mm \noindent {\bf 3.1. Gaussian field theory route} \par
\vskip 5 mm
For the free energy of the two-dimensional Gaussian field theory in a strip,
with Dirichlet boundary
conditions, the expansion (1.1) with $C = - \pi/24$ was derived in
ref.~3. The relation with Coulomb systems, discussed in Section 2,
gives for Coulomb systems the expansion (1.1) with $C = \pi/24$. \par

These considerations can be generalized to $d$ dimensions, for instance by
adapting a method used in
refs. 3 and 4, as follows. By considering the functional integral (1.3) as a
path integral on an
imaginary time interval $W$, one can express (1.3) in terms of the Hamiltonian
of a quantum field
theory in $d - 1$ space dimensions. This Hamiltonian is a sum of terms
$$H_{\bf k} = {1 \over 2} \left ( {\mu_d \over \beta} \Pi_{\bf k}^2 + {\beta
\over \mu_d} k^2
\phi_{\bf k}^2 \right )$$

\noindent where ${\bf k}$ labels all possible wave-vectors in a space of
dimension $d - 1$ and
$\Pi_{\bf k}$ is the canonical momentum conjugate to $\phi_{\bf k}$. Because of
the Dirichlet
boundary conditions $\phi = 0$ at ``times'' $0$ and $W$, (1.3) is expressed
in terms of $(0, 0)$ matrix elements~:
$$Z_G = \prod_{\bf k} \left < \phi_{\bf k} = 0 \left | e^{-WH_{\bf k}}
\right | \phi_{\bf k} = 0 \right > = \prod_{\bf k} \left ( { \beta k
\over 2 \pi \mu_d \sinh Wk} \right )^{1/2} \ \ \ . $$

\noindent Therefore
$$\beta f = - \int {d^{d-1}{\bf k} \over (2 \pi )^{d-1}} \ln \left (
{ \beta k \over 2 \pi \mu_d \sinh Wk} \right )^{1/2} \ \ \ .$$

\noindent For substracting off the ultraviolet divergences, it is convenient to
consider first
$$\beta {\partial f \over \partial W} = {1 \over 2} \int {d^{d-1} {\bf k}
 \over (2 \pi )^{d-1}} k\ \hbox{ctnh} \ Wk$$

\noindent and remove the $W = \infty$ value, which gives
$$\eqalign{
\beta {\partial f \over \partial W} &= \left. \beta {\partial f \over \partial
W} \right
|_{W=\infty} + {1 \over 2} \int {d^{d-1} {\bf k} \over (2 \pi )^{d-1}}
k\ (\hbox{ctnh} \ Wk - 1 ) \cr &= \left . \beta {\partial f \over
\partial W} \right |_{W=\infty} + {(d - 1) \Gamma \left ( {d \over 2}
\right ) \zeta (d) \over 2^d \pi^{{d \over 2}} W^d} \cr     }$$

Thus, $\beta f$ has the universal finite-size correction $- C(d)/W^{d-1}$ for
the Gaussian field
theory, and (from Section 2) the correction is $C(d)/W^{d-1}$ for a
Coulomb system, with $C(d)$ given by (1.2).

\vskip 5 mm
\noindent {\bf 3.2. Screening sum rule route} \par
\vskip 5 mm
For the slab geometry, a more direct derivation of (1.1) is possible. It avoids
any explicit
reference to the Gaussian field theory, and therefore does not rely on the
heuristic connexion of
Section 2. Instead, this derivation uses a sum rule which expresses the perfect
screening property
of a conductor. \par

Let us first consider a $d$-dimensional Coulomb system which fills the
half-space $y > 0$, with an
ideal conductor wall at $y = 0$. Let $\widehat{E}_y(0)$ be the microscopic
electric field at some point, say the origin $0$, on the boundary and
$\widehat{\rho}({\bf r})$ the microscopic charge density at ${\bf r} =
({\bf x}, y)$ ($y > 0$). The sum rule is $$\beta \int d{\bf r} \ y \left <
\widehat{E}_y(0) \widehat{\rho} ({\bf r}) \right >^T = - 1 \eqno(3.1)$$

\noindent where $<\cdots >^T$ means a truncated equilibrium statistical
average~: $<AB>^T = <AB> -
<A><B>$. One can derive$^{(8)}$ (3.1) by considering the response of the system
to an external
infinitesimal point dipole $p$ oriented along the $y$-axis and located at the
origin (on the Coulomb
system side). This dipole adds to the Hamiltonian an interaction term
$- p \widehat E_y(0)$, and by linear
response theory the average charge density at ${\bf r}$ changes by $\delta \rho
({\bf r}) = \beta p
\left < \widehat{E}_y(0) \ \widehat{\rho}({\bf r}) \right >^T$. Now, we
assume that the Coulomb system has good screening properties~: the induced
charge density $\delta \rho ({\bf r})$ is localized near the origin and has
a dipole moment which cancels $p$~: $$\int d{\bf r} \ y \ \delta \rho ({\bf
r}) = - p \eqno(3.2)$$

\noindent (since the dipole moment is defined with the origin  chosen on
the wall, there are no contributions of the surface charges on the ideal
conductor wall). The sum rule (3.1) follows. \par

Let us now add a second ideal conductor wall at $y = W$. The good screening
properties imply that,
if $W$ is large, the local structure near the origin is unchanged, up to
exponentially small effects. Therefore, the charge density $\delta \rho
({\bf r})$ is unchanged, its dipole moment (3.2) is unchanged (the
integration range on $y$ in (3.2) can be kept as $(0, \infty)$ since
$\delta \rho ({\bf r})$ is localized near $y = 0$), and the sum rule (3.1)
remains valid for the two-wall system. \par

This sum rule (3.1) can now be used for computing a universal correction to the
pressure on the wall
at $y = 0$. In free $d$-dimensional space, the Coulomb interaction, solution of
(2.3), is
$$\eqalign{
&\left . G_0({\bf r}, {\bf r}') = - \ln \left ( |{\bf r} - {\bf r}'|\right
/ a \right ) \ \ \ , \quad d = 2 \cr
&\left . G_0({\bf r} , {\bf r}') = 1 \right / |{\bf r} - {\bf r}' |^{d-2} \ \ \
 , \quad d > 2 \ \
 \cr }$$

\noindent ($a$ is some irrelevant length scale). With the present
boundary conditions, the solution of (2.3) becomes $$G({\bf r}, {\bf
r}') = G_0({\bf r}- {\bf r}') - G_0({\bf r}^* - {\bf r}') +
\sum_{n=1}^{\infty} \left [ G_0({\bf r} + n2W {\bf u} - {\bf r}')
\right .$$ $$\left. - G_0({\bf r}^* + n2W {\bf u} - {\bf r}') +
G_0({\bf r} - n2W {\bf u} - {\bf r}') - G_0({\bf r}^* - n2W {\bf u} -
{\bf r}') \right ] \eqno(3.3)$$

\noindent where ${\bf r}^* = ({\bf x}, - y)$ is the image of ${\bf r} = ({\bf
x}, y)$ with respect
to the wall at $y = 0$ and where ${\bf u}$ is the unit vector along the $y$
axis. Using
$$\left . \widehat{E}_y(0) = - \int d{\bf r} {\partial G({\bf r}, {\bf r}')
\over \partial y'}
\right |_{{\bf r}' = 0} \widehat{\rho}({\bf r})$$

\noindent we can write

$$\left . \left < \widehat{E}_y(0)^2 \right >^T = - \int d{\bf r} {\partial
G({\bf r}, {\bf r}') \over
\partial y'}\right |_{{\bf r}' = 0} \left < \widehat{E}_y(0)
\widehat{\rho}({\bf r}) \right >^T \ \
\ . $$

\noindent Since $\left < \widehat{E}_y(0) \widehat{\rho}({\bf r}) \right >^T$
is short-ranged, we
can expand $\partial G/\partial y'$ in powers of $W^{-1}$, at fixed
${\bf r}$, with the result

$$\left < \widehat{E}_y(0)^2 \right >^T = \left < \widehat{E}_y(0)^2 \right
>^T_{W=\infty} +
{\varepsilon (d - 2) (d - 1) \zeta (d) \over 2^{d-2} W^d} \int d{\bf
r}\ y \left < \widehat{E}_y(0) \widehat{\rho}({\bf r}) \right >^T$$  $$
+ O\left ( 1/W^{d+1} \right ) \eqno(3.4)$$

\noindent where $\varepsilon (d - 2) = d - 2$ if $d > 2$ and $\varepsilon (d -
2) = 1$ if $d = 2$.
Let us describe the short-range interaction between the wall at $y = 0$ and
the particles by some potential $V(y)$, and let $n(y)$ be the particle
number density. The pressure is
$$p = - \int_0^{\infty} {dV \over dy} n(y)
dy - {1 \over 2 \mu_d} \left < \widehat{E}_y(0)^2 \right > \eqno(3.5)$$

\noindent where the last term is the electrostatic pressure [if the short-range
interaction is an
infinite barrier at $y = y_0$, $-dV/dy$ must be replaced by $\beta^{-1}
\delta (y - y_0)$]. Since the screening effect makes the density profiles
near the wall $y = 0$ independent of $W$ (up to exponentially small
corrections), the related average field $$\left < \widehat{E}_y(0) \right
> = - \mu_d \int_0^{\infty} \left < \widehat{\rho}(y) \right > dy$$

\noindent is also independent of $W$. Thus, the pressure (3.5), where $\left <
\widehat{E}_y(0)^2 \right
>$ can be rewritten as $\left < \widehat{E}_y(0)^2 \right >^T + \left <
\widehat{E}_y(0)\right
>^2$, depends on $W$ only through (3.4). Using the sum rule (3.1) in (3.4)
gives for the pressure
$p$
$$\beta p = \beta p (W = \infty ) + {(d - 1) \Gamma \left ( {d \over 2} \right
) \zeta (d) \over 2^d
\pi^{{d \over 2}} W^d} + O\left ( {1 \over W^{d+1}} \right )  \ \ \ .
\eqno(3.6)$$

Since $\omega$, the grand potential per unit area, has the derivative
$${\partial \omega \over \partial W} = - p$$

\noindent the grand-potential expansion is indeed given by (1.1) and (1.2).
\par \vskip 5 mm
\noindent {\bf 3.3. The two-dimensional two-component plasma} \par \vskip 5 mm

This is a two-dimensional system of particles of charges $\pm q$. The model is
exactly
solvable$^{(9, 10)}$ at a special temperature $\beta^{-1}$ such that $\beta q^2
= 2$. For $\beta q^2
\geq 2$, a point-particle model is unstable against the collapse of pairs
of oppositely charged particles~; we work with almost-point particles,
introducing some short-distance cutoff in the interaction. In presence
of ideal conductor walls, for $\beta q^2 \geq 1$ the model would also be
unstable against the collapse of a particle onto its electric image,
unless some short-distance cutoff is also introduced in the wall-particle
attraction. \par

In two dimensions, it is convenient to represent a position ${\bf r} = (x, y)$
by the complex
coordinate $z = x + iy$, and the Coulomb interaction (3.3) can be
explicitly summed into
$$
G({\bf r},{\bf r}')=-\ln \left |{\sinh\ k(z-z')\over
\sinh\ k(z-\bar{z}')}\right | \eqno(3.7)
$$

\noindent where $k = \pi/2W$ (this result can also be obtained by a
conformal transformation from the half-plane). The calculation for the
present strip geometry is very similar to the one which has already been
done$^{(11)}$ for the half-plane case with only one ideal conductor wall.
We start with a lattice model to avoid the collapse of positive and
negative particles. There are two interwoven  lattices  $L_+$ and $L_-$
with $N_+$ and $N_-$ sites. The complex coordinates of the sites of $L_+$
(resp. $L_-$) are $\{ u_i\}_{1\leq  i\leq N_+}$
(resp. $\{ v_i\}_{1\leq  i\leq N_-}$). Positive particles can be on
the sites of $L_+$ and negative ones on those of $L_-$. Each site
contains at most one particle. We work in the grand canonical ensemble
and denote the fugacity by $\zeta$. The interaction between two
particles is given by (3.7), and furthermore each particle has a
self-interaction\footnote{$^{3}$}{In equation (2.3c) of ref. 11,
exponents 1/2 are missing for the first two factors on the r.h.s.} $-
(q^2/2) \ln |ak/\sinh k(z - \bar{z})|$. Writing

$$\sinh k(z - z') = {1 \over 2} e^{-k(z + z')} \left ( e^{2kz} - e^{2kz'}
 \right )$$

\noindent and using $\exp (2ku)$ and $\exp (2kv)$ instead of $u$ and $v$ in the
Cauchy double
alternant formula, we can follow the same steps as in ref.11 and we
obtain for the grand partition function $\Xi = \det (1 + K)$ where $K$
is the matrix $$
 K=\pmatrix{ \left({iak\zeta \over\sinh
k(u_i-\bar{u}_j)}\right)_{1\leq i\leq N_+\atop  1\leq j\leq
    N_+} & \left({iak\zeta\over \sinh k(u_i-v_j)}\right)_{
1\leq i\leq N_+\atop 1\leq j\leq N_-} \cr \left({-iak\zeta
\over \sinh k(\bar{v}_i-\bar{u}_j)}\right)_{ 1\leq i\leq
N_-\atop 1\leq j \leq N_+} & \left({-iak\zeta \over \sinh
k(\bar{v}_i-v_j)}\right)_{
    1\leq
    i\leq N_-\atop 1\leq j\leq N_-}\cr}
$$

\noindent We now consider the
continuum limit where the lattice spacing goes to zero. We define the
rescaled fugacity $m=2\pi a\zeta /S$ where $S$ is the area of a lattice
cell. The eigenfunctions $(\psi,\chi)$ of $m^{-1}K$ and its eigenvalues
$1/\lambda$ are defined by the two coupled integral equations
$$
\eqalignno{
{i\lambda\over 4W}\int_D d^2{\bf r}'({1\over\sinh
k(z-\bar{z}')}\psi({\bf r}') &+{1\over \sinh k(z-z')}\chi({\bf r}'))=
\psi({\bf r}) &(3.8a)\cr
  {-i\lambda\over
4W}\int_D d^2{\bf r}'({1\over\sinh k(\bar{z}-\bar{z}')}\psi
({\bf r}') & +{1\over\sinh k(\bar{z}-z')}\chi({\bf r}'))
  =  \chi({\bf r}) &(3.8b)\cr}
$$
\noindent where the domain of integration $D$ is the strip. In terms of
these eigenvalues,
$$\ln \Xi = \sum_{\lambda} \ln \left ( 1 + {m
\over \lambda} \right ) \ \ \ . \eqno(3.9)$$

By using the equality
$$
{\partial\over \partial \bar z} {1\over\sinh k(z-z')}=
{\pi\over k} \delta
({\bf r}-{\bf r}') , \qquad{\bf r}, {\bf r'} \in  D
$$
we can transform the integral equations (3.7) into the differential
equations
$$
2 {\partial\over \partial \bar z} \psi=\lambda i\chi \qquad \hbox{and}
\qquad 2{\partial\over \partial \bar z} i\chi=\lambda \psi
$$
In terms of the Dirac operator ${\bfsigma . \bfnabla}$, these
equations have the familiar$^{(10)}$ form
 $$({\bfsigma . \bfnabla}) \Psi =
\lambda \Psi \eqno (3.10)
$$
where here $\Psi$ is the spinor $(\psi, i\chi)$. These equations can be
combined into the Laplacian eigenvalue problem
$$
\Delta \psi=\lambda^2\psi \eqno (3.11)
$$
with boundary conditions given by the
integral equations (3.7):
$$
\psi(x,0)=-\chi(x,0) \qquad \hbox{and} \qquad \psi(x,W)=\chi(x,W)
$$
(incidentally, with such boundary conditions, $-\Delta$ turns out to
be non-hermitian).
We look for a solution of the form
$$
\psi(x,y)=e^{ivx}(Ae^{ily}+Be^{-ily}) \qquad \hbox{with} \qquad
l=i\sqrt{v^2+\lambda^2} \eqno (3.12)
$$
The boundary conditions give two linear homogeneous  equations for the
coefficients $A$ and $B$. The determinant of this system must be zero
for a non-vanishing solution to exist. This gives the relation between
$v$ and $\lambda$: $$
\cosh
(W\sqrt{v^2+\lambda^2})-\lambda{\sinh(W\sqrt{v^2+\lambda^2})\over
\sqrt{v^2+\lambda^2}}=0 \eqno (3.13)
$$ Let us define the entire fonction
$$
f(z)={1\over\cosh vW}\left[ \cosh (W\sqrt{v^2+z^2})
-z{\sinh(W\sqrt{v^2+z^2})\over\sqrt{v^2+z^2}}\right]
$$
The solutions of (3.13) are the zeros of $f$ and we have $f(0)=1$, so
$$f(z)=\prod_{\lambda\in f^{-1}(0)}(1-{z\over \lambda})$$

\noindent Therefore, for each value of $v$, a partial summation of
(3.9) can be made, giving for the grand potential $\omega$ per unit
length
$$
\beta \omega = -{1\over 2\pi}\int_{-\infty}^{+\infty}dv \ln
\prod_{\lambda \in f^{-1}(0)}(1+{m\over\lambda})
=-
{1\over \pi}\int_{0}^{+\infty}dv \ln f(-m)
$$
Now, in the limit $W\rightarrow\infty$ ,
$$
\ln f(-m)\sim
W(\sqrt{v^2+m^2}-|v|)+\ln(1+{m\over \sqrt{v^2+m^2}})-\ln(1+e^{-2|v|W})
+  O\left(e^{-mW}\right)
$$
The first term gives the bulk contribution to the grand potential, the
second the surface term and the third the universal finite-size correction. The
bulk and surface terms  would diverge if we did not take into account
the particle-particle and particle-wall short-range repulsion. We can
do so, for a repulsion of range $\sigma$, by introducing a cutoff
$v_{max} = 1/\sigma$. \par

The final result for the grand potential per unit length can be written as
$$\beta \omega = - \beta p_b W + 2 \beta \gamma_c + {\pi \over 24W} +
 O (e^{-mW})
\eqno(3.14)$$

\noindent where
$$\beta p_b = {m^2 \over 2 \pi} \left ( \ln
 {2 \over m \sigma} + 1 \right ) \eqno(3.15)$$
\noindent and
$$\beta \gamma_c = - {m \over 2 \pi} \left ( \ln {2 \over m \sigma} +
 1
 - {\pi \over 2} \right )  \eqno(3.16)$$

\noindent The bulk pressure $p_b$ is essentially the same as the one found for
an infinite
system$^{(10)}$, up to a slightly different cutoff prescription. The
surface tension $\gamma_c$ for ideal conductor walls is given by
(3.16). And, finally, we find the finite-size correction $C(2)/W = \pi
/24 W$, as expected. \par

In the argument of Section 3.2, it was assumed that the local
structure near one wall is unchanged, up to exponentially small
corrections, by the presence of another wall at a large distance $W$.
This assumption can be explicitly verified on the present model, by a
calculation of the one-body and many-body densities, by the same
method as in some previous work$^{(10, 12)}$. These densities can be
expressed in terms of Green functions $g_{ss'}({\bf r}, {\bf r}')$ ($s,
s' = \pm 1$). The one-body density of particles of sign $s$ is
$n_s({\bf r}) = m g_{ss}({\bf r}, {\bf r})$, the two-body truncated
densities are $n_{ss'}^{(2)T}({\bf r}, {\bf r}') = - m^2 g_{ss'}({\bf
r}, {\bf r}')g_{s's}({\bf r}', {\bf r})$, etc... The matrix $g$ is
defined by $g = m^{-1}K(1 + K)^{-1}$, and therefore $g_{++}$ and
$g_{-+}$ satisfy the integral equations
$$
\displaylines{
\qquad g_{++}({\bf r}_1,{\bf r}_2)+{im\over 4W} \int_{D}d^2{\bf r} \big(
{1\over\sinh k(z_1-\bar{z})}g_{++}({\bf r},{\bf r}_2) + {1\over\sinh
k(z_1-z)}g_{-+}({\bf r},{\bf r}_2)\big) \hfill\cr
\hfill = {i\over 4W\sinh
k(z_1 - \bar{z}_2)}\qquad \cr}
$$
$$\displaylines{
\qquad g_{-+}({\bf r}_1,{\bf r}_2)-{im\over 4W}
\int_{D} d^2{\bf r} (
{1\over\sinh k(\bar{z}_1-\bar{z})}g_{++}({\bf r},{\bf r}_2) +
{1\over\sinh
k(\bar{z}_1-z)}g_{-+}({\bf r},{\bf r}_2))\hfill \cr
 \hfill ={-i\over4W\sinh k(\bar{z}_1 - \bar{z}_2)}\qquad \cr}
$$

\noindent (similar equations hold for $g_{--}$ and $g_{+-}$).
Differentiating these equations with respect to $\bar{z}_1$ and $z_1$
respectively, we find differential equations which can be combined
into\footnote{$^{4}$}{Refs. 10 and 11 use slightly different
representations of the operator $K$. In the present section,
following ref.11 rather than ref. 10, we have in eq. (3.10) a spinor
$\Psi = (\psi, i\chi)$ rather than $(\psi, \chi)$ and now a  $g_{-+}$
which is $i$ times the $g_{-+}$ of ref. 10. This has no incidence on the
physical quantities.}
$$\left ( m^2 - \Delta_1 \right ) g_{++}({\bf
r}_1, {\bf r}_2) = m \delta ({\bf r}_1 - {\bf r}_2)$$
 $$g_{-+}({\bf
r}_1, {\bf r}_2) = {i \over m} \left ( {\partial \over \partial x_1} +
i {\partial \over \partial y_1} \right ) g_{++}({\bf r}_1, {\bf r}_2)
$$
Again the boundary conditions are given by the integral equations :
$$
g_{++}(x_1,0;{\bf r}_2)=-g_{-+}(x_1,0;{\bf r}_2) \qquad
\hbox{and} \qquad
g_{++}(x_1,W;{\bf r}_2)=g_{-+}(x_1,W;{\bf r}_2)
$$
For the present geometry it is useful to work with the Fourier
transforms defined by
$$
g({\bf r}_1,{\bf r}_2)=\int_{-\infty}^{+\infty}
{dl\over 2\pi}\hat{g}(y_1,y_2,l)e^{il(x_1-x_2)}
$$
The solution for $\hat{g}_{++}$ is:
$$
\displaylines
{\hat{g}_{++}(y_1,y_2,l) ={m\over 2\kappa}
e^{-\kappa |y_1-y_2|} +\cr
\big [{l\over 2\kappa}\left(
(\kappa+l)e^{\kappa (y_1+y_2)-\kappa W}-
(\kappa-l)e^{-\kappa (y_1+y_2)+ \kappa W}
 \right)
-{m\over\kappa} (\kappa-m)e^{-\kappa W}\cosh \kappa (y_1-y_2)\big ]\cr
\times \big [(\kappa-m)e^{-\kappa W}+(\kappa+m)e^{\kappa W}\big
]^{-1}\cr} $$
with $\kappa = \sqrt{l^2+m^2}$.
As $W\rightarrow +\infty$ for fixed values of $y_1$ and $y_2$, one
finds after some manipulations
 $$
\hat{g}_{++}(y_1,y_2,l)=
{m\over 2\kappa}\left(
\exp [-\kappa |y_1-y_2|] +
{\kappa-l-m\over \kappa+l+m}\exp [-\kappa (y_1+y_2)]
\right)+  O (e^{-mW})
$$
Up to an exponentially small term, this expression is the same one as
in the case of one ideal conductor wall (see eq. (3.24) of ref.10).
There are no algebraic correction in $1/W$.  This confirms
 our basic assumption of Section  3.2 that the correlation
functions are unchanged, up to exponentially small effects, by the
presence of the second wall.

\par \vskip 5 mm
\noindent {\bf 3.4. The two-dimensional one-component plasma} \par \vskip 5 mm
This is a Coulomb system with only one species of mobile particles, of
charge $q$, embedded in a background carrying a uniform charge density
of the opposite sign. In two dimensions, the model is exactly
solvable$^{(13, 14)}$ when $\beta q^2 = 2$. The half-plane geometry,
with only one ideal conductor wall, has already been studied$^{(15)}$.
Since the particles repel each other, there is no pair collapse and one
can deal with point-particles. In presence of ideal conductor walls
however, one should prevent the collapse of a particle onto its
electric image~; here we shall assume that an infinite potential
barrier keeps the particles away from the walls by a distance
$\varepsilon$, while the background extends up to the walls (this is
slightly different from what has been assumed in ref. 15). \par

For the present strip geometry, the simplest approach is to take an
appropriate limit of the two-component case. Instead of one fugacity,
one can introduce different fugacities for positive and negative
particles, and make the negative particle fugacity go to zero.
Furthermore, a uniform background of charge density $-q\eta$ generates
an electric potential $-\pi q\eta y(W - y)$ (it has the correct
Laplacian $2\pi q \eta$ and it vanishes on the walls at $y = 0$ and $y
= W$)~; this potential can be taken into account by replacing the
constant fugacity $\zeta$ of the positive particles by a
position-dependent fugacity $\zeta \exp [2 \pi \eta y(W - y)]$ (here
$\beta q^2 = 2$), and adding to the grand potential the background
self-energy which is $(1/6) \beta^{-1} \pi \eta^2 W^3$ per unit length.
Then, the set of equations (3.8) reduces to one equation $${i \lambda
\over 4W} \int_D d^2{\bf r}' {e^{2\pi \eta y'(W-y')} \over \sinh {\pi
\over 2W} (z - \bar{z}')} \psi ({\bf r}') = \psi({\bf r})
\eqno(3.17)$$

\noindent where $D$ now is the strip between $y = \varepsilon$ and $y = W -
\varepsilon$. \par

Since $z - \bar{z}'$ does not vanish if ${\bf r}, {\bf r}' \in D$,
(3.17) indicates that $\psi ({\bf r})$ is an analytical function of $z$.
Because of the translational invariance along the $x$-axis, $\psi$ must
depend on $x$ through a factor $\exp (ivx)$, and therefore $\psi = \exp
(ivz)$. Using this $\psi$ in (3.17) and performing the integral upon
$x'$ gives the relation between $v$ and $\lambda$  $$
{\lambda\over e^{2Wv}+1}
\int_{\epsilon}^{W-\epsilon} e^{2[\pi\eta y'(W-y')+vy']} dy'
=1
$$

\noindent Using this $\lambda$ in (3.9) (with the background self-energy
added), performing the
integral upon $y'$, and making the change of variable $v = 2 \pi \eta Wt$,
gives for the grand
potential
$$
\eqalign{-\beta\omega & = -{\pi\eta^2\over 6}W^3 \cr
 & +  2\eta W \int_{0}^{+\infty}dt\ln  \Big\{ 1+ {m\over
2\sqrt{2\eta}} {e^{{\pi\over 2}W^2\eta (1+4t^2)}\over 2\cosh (2\pi\eta
W^2 t)} [\hbox{erf}\ (\sqrt{2\pi\eta}W({1\over 2}-t-{\epsilon\over
W})) \cr
&+ \hbox{erf}\ \big(\sqrt{2\pi\eta}W({1\over
2}+t-{\epsilon\over W}) \big)
\big ]\Big \}\cr}
$$
where erf is the error function. Now we  split the integral over
$t$ into an integral going from $0$ to $1/2$ and another one going from
$1/2$ to $+\infty$. We write $$
-{\pi\eta^2 W^3\over 6}=2\eta W\int_{0}^{1/2}
\ln e^{-\pi\eta W^2 (1-2t)^2/2}dt
$$
and add this term to the first integral. The resulting integral from
$0$ to $1/2$ can be separated in three terms which give, in the limit
$W\rightarrow +\infty$, up to exponentially small terms, the bulk
pressure, a first contribution for the surface tension and the
universal finite-size correction
$$
\eqalignno{
\beta p_b W & =  2W\eta\int_{0}^{1/2} dt
\ln{m\over \sqrt{2\eta}} =\eta W \ln{m\over \sqrt{2\eta}}&(3.18)
\cr
-2\beta\gamma_c^{(1)} & =  2W\eta\int_{0}^{1/2}dt \ln \Big[
{\sqrt{2\eta}\over m}e^{-{\pi\over 2}\eta W^2(1-2t)^2}
+\cr
&{
\hbox{erf}\ (\sqrt{2\pi\eta}W({1\over 2}-t-{\epsilon\over
W}))+ \hbox{erf}\ (\sqrt{2\pi\eta}W({1\over
2}+t-{\epsilon\over W})) \over 2} \Big ] dt  \cr
  =  \sqrt{{2\eta\over \pi}}&\int_{0}^{+\infty}dt
\ln[e^{-t^2}{\sqrt{2\eta}\over m}+{1\over 2}(1+\hbox{erf}\
(t-\sqrt{2\pi\eta}\epsilon))] dt + O(e^{-mW}) &(3.19)
\cr
-{\pi\over 24W} & =  -2W\eta\int_{0}^{+\infty}dt \ln(1+e^{-4\pi\eta
W^2t})+ O(e^{-mW}) \cr}
$$
The  integral from $1/2$ to $+\infty$ contributes only to the surface
tension; it can be reexpressed as
$$
-2\beta\gamma_c^{(2)}=
\sqrt{{2\eta\over\pi}}\int_{0}^{+\infty}dt
\ln\left[1+{m\over
2\sqrt{2\eta}}e^{t^2}(1-\hbox{erf}\ (t+\epsilon\sqrt{2\pi\eta}))
\right]+ O (e^{-mW}) \eqno (3.20)
$$
The final result for the grand potential is
$$
\beta \omega=
-\beta p_b W + 2\beta (\gamma_c^{(1)}+\gamma_c^{(2)})+{\pi\over 24W}
+ O (e^{-mW})
$$

\par \vskip 5 mm
\noindent {\bf 3.5. Strip with plain hard walls} \par \vskip 5 mm

The solvable two-dimensional models can also be used for studying the case of a
strip with plain
hard walls. The walls confine the particles, but there is no Dirichlet
boundary condition for the
electric potential, which freely ``leaks out''. The Coulomb
interaction is just the free space one $- \ln (|{\bf r} - {\bf
r}'| /a)$. In that case, there is no finite-size correction of order
$1/W$ in the grand potential or free energy per unit length of the
strip, as shown hereafter. \par

For the two-component plasma, the grand partition function is still given by
(3.9) and (3.11), but
the boundary conditions now are$^{(2)}$ \par
\item{-} on the boundary $y = 0$, $\psi = g_0$, $\partial \psi/\partial
 \bar{z} = \bar{h}_0$ \par
\item{-} on the boundary $y = W$, $\psi = g_W$, $\partial \psi/\partial
\bar{z} = \bar{h}_W$ \par

\noindent where $g_0$, $h_0$ (resp. $g_W$, $h_W$) are analytical functions in
the domain $y < 0$
(resp. $y > W$) vanishing at infinity. For $\psi$ of the form (3.12),
these boundary conditions become
$$\psi (x, 0) = 0 \quad , \qquad {\partial \psi \over \partial \bar{z}} (x, W)
= 0 \qquad (v > 0)$$
$$\psi (x, W) = 0 \quad , \qquad {\partial \psi \over \partial \bar{z}}
 (x, 0) = 0 \qquad (v < 0) \
\ \ . $$

\noindent For each value of $v$, these equations give two linear homogeneous
equations for the
coefficients $A$ and $B$, the compatibility of which leads to the relation

$$\cosh (W \sqrt{v^2 + \lambda^2}) + |v| {\sinh (W \sqrt{v^2 + \lambda^2})
\over \sqrt{v^2 +
\lambda^2}} = 0$$

\noindent instead of (3.13). Now
$$f(z) = e^{-W|v|} \left [ \cosh (W \sqrt{v^2 + z^2}) + |v| {\sinh (W
\sqrt{v^2 + z^2}) \over \sqrt{v^2 + z^2}} \right ]$$

\noindent and, in the limit $W \to \infty$,
$$\ln f(-m) \sim W \left ( \sqrt{v^2 + m^2}) - |v| \right ) + \ln
\left [ {1 \over 2} \left ( 1 + {|v| \over \sqrt{v^2 + m^2}} \right )
\right ] +  O \left ( e^{-mW} \right )$$

\noindent which gives for the grand potential $\omega$ per unit length
$$\beta \omega = - \beta p_b W + 2 \beta \gamma + O \left (
e^{-mW} \right ) $$

\noindent The bulk pressure $p_b$ is the same as in (3.15), as expected.
The surface tension $\gamma$ is now given by
$$\beta \gamma = {m \over 2 \pi} \left ( {\pi \over 2} - 1 \right )$$

\noindent (an already known result$^{(10)}$). And there is no finite-size
correction, except for
an exponentially small one. \par

For the one-component plasma in a strip with plain hard walls, the canonical
free energy has
been computed in ref. 16. In our notation, with the de Broglie
wavelength and the length $a$ in the logarithmic potential taken as
unity, the result for the free energy $f$ per unit length is

$$\beta f = W {\eta \over 2} \ln {\eta \over 2 \pi^2} - 2 \sqrt{{\eta
\over 2 \pi}} \int_0^Y dt \ \ln {\hbox{erf} (Y + t) + \hbox{erf} (Y - t)
\over 2}$$

\noindent where $Y = W (\pi \eta /2)^{1/2}$ and erf is the error
function. In the large-$W$ limit one finds
$$\beta f \sim W {\eta \over 2} \ln {\eta \over 2\pi^2} - 2 \sqrt{{\eta
\over 2 \pi}} \int_0^{\infty} dt \ \ln {1 + \hbox{erf}(t) \over 2} +
{e^{-\pi \eta W^2} \over 4\pi^2  \eta^{1/2} W^2} + \cdots$$

\noindent The first term corresponds to the bulk free energy, the second one to
the surface
tensions$^{(17)}$ on both boundaries, and the next term is indeed more than
exponentially
small~; there is no algebraic finite-size correction. \par

\vskip
5mm \noindent {\bf 4. DISK} \vskip 5mm
This Section is about a two-dimensional case~: a Coulomb system in a disk of
radius $R$, with an
ideal conductor boundary. We derive the expansion (1.4) in general, and
check it on two solvable models. \par \vskip 5 mm

\noindent {\bf 4.1. General derivation} \vskip 5 mm
We have not been able to make a direct derivation of (1.4) using, for
instance, screening properties of the Coulomb system. We can only rely
on the heuristic method of Section 2, leading to the relation (2.5)
between the Coulomb system and the Gaussian field theory. For the
latter, it has been shown$^{(5, 6, 18)}$ that $-\ln Z_G$ has a
universal finite-size term $-(1/6)\ln R$. From (2.5), we obtain
(1.4). \par \vskip 5 mm

\noindent {\bf 4.2. The two-component plasma} \vskip 5 mm

The two-component plasma model, at $\Gamma = 2$, is also solvable in the
present geometry. In
terms of complex coordinates $z = r e^{i\theta}$, with the origin at the
center $O$ of
 the disk, the Coulomb potential
solution of (2.3) with Dirichlet boundary conditions is obtained by the
method of images as
$$G({\bf r}, {\bf r}') = - \ln\left | {R(z -
z') \over z \bar{z}' - R^2} \right | $$

\noindent Like in Section 3.3, we obtain (3.9) for the grand partition
function. Now the sum is on the eigenvalues $\lambda$ defined by the
two coupled integral equations $$\eqalignno{
&{\lambda \over 2 \pi} \int_D d^2 {\bf r}' \left ( {-R \over z \bar{z}' - R^2}
\psi ({\bf r}') + {1
\over z - z'} \chi ({\bf r}') \right ) = \psi ({\bf r})  &(4.1a) \cr
&{\lambda \over 2 \pi} \int_D d^2 {\bf r}' \left ( {1
\over \bar{z} - \bar {z}'} \psi ({\bf r}') + {-R \over  \bar{z}z' - R^2}
\chi ({\bf r}') \right ) = \chi ({\bf r})  &(4.1b) \cr }$$

\noindent where the domain of integration $D$ is the disk. Using the equality
$${\partial \over \partial \bar{z}} \ {1 \over z - z'} = \pi \delta ({\bf r} -
{\bf r}')$$

\noindent we again find the Laplacian equation (3.11)  with now the
boundary condition $\chi = e^{i\theta} \psi$, i.e.
$$\left . {1 \over \lambda} e^{i \theta} \left ( {\partial \over
\partial r} + {i \over r} \ {\partial \over \partial \theta} \right )
\psi (r, \theta ) - e^{i\theta} \psi (r, \theta ) \right |_{r= R} = 0
\eqno(4.2)$$

\noindent (again, $- \Delta$ is non-hermitian with that boundary
condition). \par
\def\Z{ {\rm Z \kern -.27cm \angle \kern .02cm}}
Because of the circular symmetry, we look for eigenfunctions of the form
$$\psi = I_{\ell} (\lambda r) e^{i\ell \theta} \qquad , \quad \ell \in
\Z$$

\noindent where $I_{\ell}$ is a modified Bessel function. The boundary
condition (4.2) imposes
$$I'_{\ell}(\lambda R) - {\ell \over \lambda R}
I_{\ell} (\lambda R) - I_{\ell} (\lambda R) = 0$$ \noindent i.e.
$$I_{\ell + 1} (\lambda R) - I_{\ell} (\lambda R) = 0 \ \ \ . \eqno(4.3)$$

\noindent All the eigenvalues $\lambda$ are obtained by taking,
for each value of $\ell \in \Z$, all
the non-zero roots of (4.3). Since (4.3) is invariant under the
transformation
 $\ell \rightarrow - \ell - 1$, it is
enough to consider the eigenvalues $\lambda_{\ell}$ associated to $\ell \in
I\hskip - 1 mm N$,
counting each of them twice. Then, as in Section 3.3,  for each value of
$\ell$ a partial summation of (3.9) can be made by noting that the
corrresponding eigenvalues $\lambda_{\ell}$ are the zeros of the
function $f_{\ell}(z)$~: $$f_{\ell}(z) = \ell ! \left ( {2 \over zR}
\right )^{\ell} \left [ I_{\ell} (zR) - I_{\ell + 1}(zR) \right ],
\qquad \ell \geq 0 \ \ \  \eqno(4.4)$$

\noindent The prefactor in (4.4) has been chosen in such a way that
$f_{\ell}(0) = 1$. Therefore,
the entire function $f_{\ell}(z)$ can be written as the product
$$
f_{\ell}(z) = \prod_{\lambda_{\ell}} \big (1 - {z \over
\lambda_{\ell}}\big )$$

\noindent and
$$\sum_{\lambda_{\ell}} \ln \left ( 1 + {m \over \lambda_{\ell}} \right
) = \ln \prod_{\lambda_{\ell}} \left ( 1 + {m \over \lambda_{\ell}}
\right ) = \ln f_{\ell}(-m)$$

\noindent making (3.9) a sum on $\ell$ only~:
$$\ln \Xi = 2 \sum_{\ell = 0}^{\infty} \ln \left \{ \ell ! \left ( {2
\over mR} \right )^{\ell} \left [ I_{\ell} (mR) + I_{\ell + 1} (mR)
\right ] \right \} \ \ \ . \eqno(4.5)$$

For obtaining a large-$R$ asymptotic expansion of (4.5), one can use
the Debye expansion$^{(19)}$
of $I_{\ell}(mR)$. This expansion is the one which is appropriate here, since
both $mR$ and
$\ell$ become large. It is convenient to split (4.5) as
$$\ln \Xi = S_1 + S_2$$

\noindent where
$$S_1 = 2 \sum_{\ell = 0}^{\infty} \ln \left [ \ell ! \left ( {2 \over
mR} \right )^{\ell} I_{\ell} (mR) \right ]$$

\noindent and
$$S_2 = 2 \sum_{\ell = 0}^{\infty} \ln \left [ 1 + {I_{\ell + 1}(mR)
\over I_{\ell}(mR)} \right ]$$

\noindent since $S_1$ is $\ln \Xi$ for a plain hard wall boundary and
has already been evaluated in ref. 2. As to $S_2$, rewritten and cut
off as $$S_2 = 2 \sum_{\ell = 0}^L \ln \left [ 1 + {I'_{\ell}(mR)
\over I_{\ell}(mR)} - {\ell \over mR} \right ]$$

\noindent it can be evaluated by similar techniques, using the Debye expansion
of $I_{\ell}$ and
$I'_{\ell}$, and using for the sum on $\ell$  the Euler-MacLaurin
summation formula . If we now take into
account the particle-particle and particle-wall short range repulsions
by choosing the cutoff as $L = R/\sigma$, we find $$S_2 = mR \ln {2
\over m \sigma} + {1 \over 2} \ln 2 - {\pi \over 4} + O \left ( {1
\over mR} \right ) \ \ \ . \eqno(4.6)$$

\noindent In (4.6), all terms which vanish as the cutoff $\to
\infty$ have already been removed.
\par

Adding $S_1$ (from ref. 2) and $S_2$, one finds for the grand potential
$\Omega$, in the case of an
ideal conductor boundary,
$$\beta \Omega = - \ln \Xi = - \beta p_b \pi R^2
+ \beta \gamma_c 2 \pi R + {1 \over 6} \ln \
mR - {5 \over 12} - {1 \over 6} \ln 2 - 2 \zeta ' (-1) + {\pi \over
4} \eqno(4.7)$$

\noindent where the bulk pressure $p_b$ and the surface tension $\gamma_c$ are,
as expected, given
by the same expressions (3.15) and (3.16) as in the strip geometry. And
(4.7) has the expected universal finite-size correction $(1/6) \ell n
R$.

One should note that there is no $\ln R$ term in $S_2$~; it
comes entirely from $S_1$. In other words, the universal finite-size
correction is the same one for a plain hard wall and for an ideal
conductor wall, unlike in the strip geometry. \par \vskip 5 mm
\vfill\eject
\noindent {\bf 4.3. The one-component plasma} \vskip 5 mm

We consider the same model as in Section  3.4, but in the disk geometry.
The electric potential generated by the background is
$-{1\over 2}\pi\eta q(R^2-r^2)$. The background self-energy adds to the
grand potential a term ${1\over 4}\beta^{-1}(\pi\eta R^2)^2$ The
 eigenvalue integral equation  now is
$$
{-\lambda R\over 2\pi}
\int_{D}
{\exp[\pi\eta(R^2-r^2)]\over
\bar{z}'z-R^2}\psi({\bf r}')d^2{\bf r}'= \psi({\bf r})\eqno (4.8)
$$
Again we notice that $\psi$ is analytical. Because of the rotational
invariance, $\psi$ must depend on $\theta$ through a factor
$\exp(i\ell\theta)$, and therefore $\psi({\bf r})=z^\ell$ with $\ell$ a
non-negative integer. Using this form in (4.8) we have the
relation between $\lambda$ and $\ell$
$$
\lambda\exp[\pi\eta R^2]\int_{0}^{R-\epsilon}{
\left({r\over R}\right)^{2\ell+1}\exp[-\pi\eta r^2]
dr} = 1
$$
The grand potential becomes, after some manipulations,
$$
\eqalignno{
\beta\Omega & =  {1\over 4}(\pi\eta R^2)^2 \cr
 & -  \sum_{\ell=0}^{+\infty}
\ln\left[
1+{mR\exp(\pi\eta R^2)\over 2(\pi\eta R^2)^{\ell+1}}
\gamma(\ell+1,\pi\eta(R-\epsilon)^2)
\right] &(4.9)\cr}
$$
where $\gamma(n,x)=\int_{0}^{x}{t^{n-1}\exp(-t)dt}$ is the incomplete
gamma function.
Let $N=\pi\eta R^2$. We split the sum over $\ell$ in eq.~(4.9) into
a sum $S_1$ for $0 \leq \ell \leq N$  and a sum $S_2$ for
$\ell>N$. Using
$$
e^N N^{-(\ell+1)} \ell ! \sim \sqrt{{2\pi\over N}} \exp \big
[{(\ell-N)^2\over 2N} + {\ell-N\over 2N}\big ] \eqno(4.10)
 $$
we have in the limit $R\rightarrow +\infty$
$$
\eqalignno{S_1 & =
(N+1)\ln({mR\over 2}) \cr
& +  \sum_{\ell=0}^{N} \big[ -(\ell+1)\ln N + N +\ln \ell! \big] \cr
 & + \sum_{\ell=0}^{N} \ln \big[{\sqrt{2\eta}\over m}
e^{-{(\ell-N)^2\over 2N}}+{1\over \ell!}\gamma(\ell+1,\pi\eta
(R-\epsilon)^2)  \big] &(4.11)\cr}
$$
Using Stirling's formula , the asymptotic formulas for
$R\rightarrow +\infty$
$$
\eqalignno{
\sum_{\ell=0}^{N}{\ln \ell!} & =  (N+1)\ln N! - {N^2\over 2}\ln N
\cr
 & +  {N^2\over 4} - {1\over 2}N\ln N - {1\over 12}\ln N +  O(1)
&(4.12) \cr}
$$
and
$$
{1\over \ell!}\gamma(\ell+1,\pi\eta (R-\epsilon)^2)={1\over 2}
\left[
1+\hbox{erf}\
\left({N-\ell\over\sqrt{2N}}-\epsilon\sqrt{2\pi\eta}\right) \right] +  O
({1\over\sqrt{N}}) \eqno (4.13) $$
we obtain, in the small -$\epsilon$ limit,
$$
\eqalignno{
S_1  &=  - {1\over 4}(\pi\eta R^2)^2 - \pi\eta R^2
\ln \left( {m\over\sqrt{2\eta}} \right) +  {1\over6} \ln \left[ (2/m)^6
(\pi\eta)^{7/2} R \right] \cr
&+ \sqrt{2\pi\eta} R
\int_0^{+\infty} dt\ln \big [{\sqrt{2\eta}\over m} e^{{-t}^2} + {1\over
2} (1 + \hbox{erf}\ (t-\epsilon \sqrt{2\pi\ eta}))\big ]  +O(1)
&(4.14)\cr}
$$
In $S_2$ we use (4.13) and (4.10) to find
$$
S_2=\sqrt{2\pi\eta}R\int_0^{+\infty} dt
\ln\left[1+{m\over 2\sqrt{2\eta}}e^{t^2}\left(
1-\hbox{erf}\ (t+\epsilon\sqrt{2\pi\eta})\right)\right] \eqno(4.15)
$$
Putting (4.14) and (4.15) in (4.9) we have
$$
\beta \Omega=
-\beta p_b\pi R^2  + 2\pi R\beta (\gamma_c^{(1)}+\gamma_c^{(2)})
+{1\over 6} \ln \left[ \left({2\over m}\right)^6
(\pi\eta)^{7/2} R \right]
+ O (1) \eqno (4.16)
$$
where the bulk pressure $p_b$ is given by (3.18)
and  $\gamma_c^{(1)}$ and
$\gamma_c^{(2)}$  are given by (3.19) and (3.20) in the
small-$\epsilon$ limit. We find the universal finite-size correction
${1\over 6}\ln R$.

 It is interesting to compare the surface tension for
the one-component plasma  in the limit $\epsilon\rightarrow 0$ and the
surface tension for the two-component plasma in the limit
$\sigma\rightarrow 0$. We have in that limit  for the one-component
plasma $$ \gamma_c^{(2)}\sim {m\over 4\pi}\ln \epsilon
$$
and for the two-component plasma
$$
\gamma_c\sim {m\over 2\pi}\ln \sigma
$$
The surface tension of the one-component plasma is one-half the one of
the two-component plasma.

\par \vskip
5mm \noindent {\bf 5. CONCLUSION} \vskip 5mm
The grand potential of a Coulomb system has universal finite-size
corrections very similar to the ones which occur for a simple
critical system : the massless Gaussian field theory. The boundary
conditions however play an important r\^ole. In the geometry of a
slab, if the boundaries are ideal conductor walls, a Coulomb system
has a grand potential with a universal finite-size correction which
is just the opposite of the one found for a Gaussian field with
Dirichlet boundary conditions; but if the boundaries are plain hard
walls, the Coulomb system no longer exhibits that finite-size
correction. In the geometry of a disk, a two-dimensional Coulomb
system has a finite-size correction to its grand potential which is
the same one for an ideal conductor wall and for the previously
studied$^{(2)}$ case of a plain hard wall (in both cases, again the
correction is opposite to the one occuring for the Gaussian field).
An argument given in ref. 2 essentially says that a Coulomb system
with plain hard walls looks somewhat like an ideal conductor when
seen from the outside; this might be relevant for explaining that
the curved boundary of a disk generates the same correction
$(1/6)\ln R$ in both plain hard wall and ideal conductor wall
cases. Nevertheless, a unified treatment of the different possible
boundary conditions still has to be found.

In the case of a slab with ideal conductor walls, the finite-size
correction to the grand potential of a Coulomb system was derived on
the sole basis of the screening effect (Section 3.2). A more
general derivation on the basis of the screening effect should be
found for other geometries and boundary conditions.

\vfill \supereject \centerline{\bf \underbar{References}} \par
  \vskip 5mm
\item{1.} J.P. Forrester, {\it J. Stat. Phys.} {\bf 63} : 491
(1991)

\item{2.} B. Jancovici, G. Manificat, and C. Pisani, {\it J. Stat.
Phys.} {\bf 76} : 307 (1994)

\item{3.} H.W.J. Bl\"ote, J.L. Cardy, and M. P. Nightingale, {\it
Phys. Rev. Lett.} {\bf 56} : 742 (1986).

\item{4.} I. Affleck, {\it Phys. Rev. Lett.} {\bf 56} : 746
(1986).

\item{5.} J.L. Cardy and I. Peschel, {\it Nucl. Phys. B} {\bf 300}
[FS 22] : 377 (1988).

 \item{6.} J.L. Cardy, in {\it Fields, Strings and Critical
Phenomna, Les Houches 1988}, E. Br\'ezin and J. Zinn-Justin, eds.
(North-Holland, Amsterdam, 1990).

\item{7.} B. Jancovici, {\it J. Stat. Phys.} {\bf 80} : \quad (1995).

\item{8.} B. Jancovici, {\it J. Phys.} (France) {\bf 47} : 389
(1986).

\item{9.} M. Gaudin, {\it J. Phys.} (France) {\bf 46} : 1027 (1985).

\item{10.} F. Cornu and B. Jancovici, {\it J. Chem. Phys.} {\bf
90} : 2444 (1989).

\item{11.} P.J. Forrester, {\it J. Chem. Phys. } {\bf 95} : 4545
(1990).

\item{12.} B. Jancovici and G. Manificat, {\it J. Stat. Phys.}
{\bf 68} : 1089 (1992).

\item{13.} A. Alastuey and B. Jancovici, {\it J. Phys. } (France)
{\bf 42} : 1 (1981).

\item {14.} B. Jancovici, {\it Phys. Rev. Lett.} {\bf 46} : 386
(1981).

\item {15.} P.J. Forrester, {\it J. Phys. A} {\bf 18} : 1419
(1985).

\item {16.} P.J. Forrester and E.R. Smith, {\it J. Phys. A} {\bf
15} : 3861 (1982).

\item {17.} V. Russier, J.P. Badiali, and M.L. Rosinberg, {\it J.
Phys. C} {\bf 18} : 707 (1985)

\item {18.} O. Alvarez, {\it Nucl. Phys. B} {\bf 216} : 125 (1983).

\item {19.} M. Abramowitz and I.A. Stegun, {\it Handbook of
Mathematical Functions} (National Bureau of Standards, Washington
D.C., 1964).

 \bye